\documentclass[11pt,a4paper]{article}
\usepackage{jheppub}

\usepackage{epsfig}
\usepackage{longtable}
\usepackage{subfigure}
\usepackage{amsfonts}

\title{Exploring the heavy quark sector of the Bestest Little Higgs model at the LHC}
\author{ Stephen Godfrey, Thomas Gr\'egoire, Pat Kalyniak, Travis A.W. Martin, Kenneth Moats}
\affiliation{Ottawa-Carleton Institute for Physics, Department of Physics, Carleton University, Ottawa, Canada K1S 5B6}
\emailAdd{godfrey@physics.carleton.ca}
\emailAdd{gregoire@physics.carleton.ca}
\emailAdd{kalyniak@physics.carleton.ca}
\emailAdd{tmartin@physics.carleton.ca}
\emailAdd{kmoats@physics.carleton.ca}

\abstract{We present discovery limits for heavy quarks in the Bestest Little Higgs model via pair production at the LHC running at $\sqrt{s} = 7~\mathrm{TeV}$. We study pair produced heavy top-like quarks decaying to $b\bar{b}W^+W^-$ and $t\bar{t}ZZ$ final states and singly produced heavy top-like quarks via t-channel $W$ exchange. These results are compared to currently available limits on heavy top-like quark cross sections from CMS (with 1.14~fb$^{-1}$ integrated luminosity) for two scenarios of Yukawa couplings. We find that the CMS data limits the mass of the lightest top partner to larger than 413~GeV in the first scenario, where the two lightest top-partners have a small mass splitting, and 364~GeV in the second scenario, where the mass splitting between the two lightest top-partners is larger.}

\begin{document}
\maketitle

\section{Introduction}

As the CMS and ATLAS experiments at the Large Hadron Collider (LHC) collect data, large regions of parameter space are being ruled out in many models of beyond the Standard Model physics.  For example, the parameter space of supersymmetric models is shrinking rapidly leading to an increasing need for fine-tuning~\cite{Chatrchyan:2011zy,Aad:2011qa,Conley:2011nn,Papucci:2011wy}.  Little Higgs (LH) models \cite{LHoriginal} on the other hand already have strong constraints from electroweak precision data. These constraints typically require the new gauge bosons of LH models to be quite heavy~\cite{Csaki:2002qg,Csaki:2003si}, but the LHC is starting to be competitively sensitive~\cite{Diener:2010sy,Collaboration:2011dca}. In most Little Higgs models, the top partners are heavier than the new gauge bosons, and this can lead to significant fine-tuning in the Higgs potential~\cite{Casas:2005ev}.

A recently proposed model called the Bestest Little Higgs ~\cite{Schmaltz:2010ac} overcomes these difficulties by including separate symmetry breaking scales at which the heavy gauge boson and top partners obtain their masses. This model features a custodial $SU(2)$ symmetry, has heavy gauge boson partner masses above the already excluded mass range, and has relatively light top partners below the upper bound from fine-tuning.

The heavy fermion sector of the Bestest Little Higgs model has a rich phenomenology, with four heavy top partners ($+2/3$ charge), a heavy bottom partner ($-1/3$ charge) and an exotic $+5/3$ charge quark, all of which could be relatively light. Discovery of a heavy top partner has been explored in a number of other scenarios~\cite{Azuelos:2004dm}. In this paper, we explore the phenomenology of the heavy fermion sector of the Bestest Little Higgs model at the LHC, including the discovery reach. With the number of heavy quarks in this model, cascade decays to non-SM particles can occur. However, these processes have a negligible effect on the results we present, since the discovery reach is dominated by the contributions from the lightest top partner.
\section{The Model}
The Bestest Little Higgs model is based on a non-linear sigma model with a global $SO(6)_A \times SO(6)_B$ that is broken by a scalar $\Sigma$ field to a diagonal $SO(6)_V$ at scale $f$. The non-linear sigma field, $\Sigma$, contains two scalar triplets ($\eta$, $\phi$), two Higgs doublets ($h_1$, $h_2$) and a heavy, real singlet ($\sigma$). Below the symmetry breaking scale $f$, the singlet is integrated out, which results in a quartic coupling for the Higgs doublets and spontaneous symmetry breaking, as with other two Higgs doublet models (such as in~\cite{Rathsman:2011yv}).

This model introduces a number of new heavy quarks, arranged into four multiplets:  $Q$, $Q^\prime_a$, $U^c$ and $U^{\prime c}_5$. The first multiplet, $Q$, is a $\bf{6}$ of $SO(6)_A$, $Q^\prime_a$ is a $\bf{2}$ of $SU(2)_A$, $U^c$ is a $\bf{6}$ of $SO(6)_B$, and $U^{\prime c}_5$ is a $\bf{1}$ of $SU(2)_{A,B}$. $SU(2)_{A,B}$ are subgroups of $SO(6)_{A,B}$, whose diagonal subgroup is the Standard Model $SU(2)_L$. 

Heavy fermion Yukawa interactions in this model take the form of:~\cite{Schmaltz:2010ac}
\begin{eqnarray}
\mathcal{L}_t &= y_1 f Q^T S \Sigma S U^c + y_2 f Q^{\prime T}_a \Sigma U^c + y_3 f Q^T \Sigma U^{\prime c}_5 + h.c., \label{eq:Lyuk}
\end{eqnarray}
where $S = \mathrm{diag}(1,1,1,1,-1,-1)$. The quark multiplets are given by
\begin{eqnarray*}
Q^T &=& (\frac{1}{\sqrt{2}}(-Q_{a1} -Q_{b2}), \frac{i}{\sqrt{2}}(Q_{a1} -Q_{b2}), \frac{1}{\sqrt{2}}(Q_{a2} -Q_{b1}), \frac{i}{\sqrt{2}}(Q_{a2} -Q_{b1}),Q_5,Q_6), \cr
Q^{\prime T}_a &=& \frac{1}{\sqrt{2}}(-Q^\prime_{a1},i Q^\prime_{a1},Q^\prime_{a2},i Q^\prime_{a2},0,0), \cr
U^{c T} &=& (\frac{1}{\sqrt{2}}(-U^c_{b1} -U^c_{a2}), \frac{i}{\sqrt{2}}(U^c_{b1} -U^c_{a2}), \frac{1}{\sqrt{2}}(U^c_{b2} -U^c_{a1}), \frac{i}{\sqrt{2}}(U^c_{b2} -U^c_{a1}),U^c_5,U^c_6), \\
U^{\prime c T}_5 &=& (0,0,0,0,U^{\prime c}_5,0). 
\end{eqnarray*}
In this Yukawa potential, the first term in Eq.~\ref{eq:Lyuk} breaks $SO(6)_A$ and $SO(6)_B$, while the second term preserves $SO(6)_B$ and the third preserves $SO(6)_A$. Collectively, all three terms break the symmetries protecting the Higgs and generate a finite contribution to the Higgs potential.

These multiplets contain eight degrees of freedom, of which two form the SM third generation left handed doublet ($q_3$) and right handed up-type singlet ($u_3^c$), which remain massless prior to electroweak symmetry breaking (EWSB):
\begin{eqnarray}
q_3 &=& \displaystyle\frac{y_2}{\sqrt{y_1^2+y_2^2}} Q_a - \displaystyle\frac{y_1}{\sqrt{y_1^2+y_2^2}} Q_a^\prime \cr\cr\cr
u_3^c &=& \displaystyle\frac{y_3}{\sqrt{y_1^2+y_3^2}} U_5^c - \displaystyle\frac{y_1}{\sqrt{y_1^2+y_3^2}} U^{\prime c}_5,
\label{eq:q3}
\end{eqnarray}
where we have assumed $y_i \in \mathbb{R}$.

Diagonalization of the fermion mass matrix results in a top quark with a mass proportional to the electroweak vacuum expectation value (vev), $v_{EW}$, and six remaining heavy quarks with masses proportional to the decay constant, $f$. The model contains four top-like quarks ($T$, $T_5$, $T_6$, $T_b^{2/3}$), one bottom-like quark ($B$), and an exotic quark ($T_b^{5/3}$) with a $+5/3$ electric charge. 

While analytic forms can be found for the masses by expanding in powers of $v_{EW}/f$, these forms lose validity for small values of the $f$ parameter. Instead, more precise values of the masses for these states can be found through numerical diagonalization of the square of the fermion mass matrix.  Additionally, in the region where $y_2 \approx y_3$, the masses of $T_5$ and $T$ are degenerate at lowest order.  Consequently, different diagonalization schemes are required for the region where $y_2 \approx y_3$ versus the region where $|y_2 - y_3| > 0$. 

We quote the order $(v_{EW}/f)^2$  analytic forms of the masses for the case $|y_2-y_3|>0$ and $f>v$ in Eq.~\ref{eq:Qmass}. Since the higher order terms are significant for small values of $f$, our numerical results do take these into account.
\begin{eqnarray}
&M_{t}^2 &= \frac{9 y_1^2 y_2^2 y_3^2 v^2 \sin^2(\beta)}{(y_1^2+y_2^2)(y_1^2+y_3^2)} \nonumber \\ \nonumber \\
&M_{T}^2 &= (y_1^2+y_2^2) f^2+\frac{9 y_1^2 y_2^2 y_3^2 v^2 \sin^2(\beta)}{(y_1^2+y_2^2)(y_2^2-y_3^2)} \nonumber\\ \nonumber\\
&M_{B}^2 &= (y_1^2+y_2^2) f^2 \label{eq:Qmass} \\ \nonumber \\
&M_{T_5}^2 &= (y_1^2+y_3^2) f^2-\frac{9 y_1^2 y_2^2 y_3^2 v^2 \sin^2(\beta)}{(y_1^2+y_3^2)(y_2^2-y_3^2)} \nonumber\\ \nonumber \\
&M_{T_6}^2 &= M_{T_b^{2/3}}^2 = M_{T_b^{5/3}}^2 =y_1^2 f^2 \nonumber
\end{eqnarray}

In Eq.~\ref{eq:Qmass}, $v^2 =v_{EW}^2 \equiv v_1^2 + v_2^2$ and $\tan\beta \equiv v_1/v_2$, where $v_1$ and $v_2$ are the vevs acquired by $h_1$ and $h_2$, respectively, through electroweak symmetry breaking.  Identifying the top quark Yukawa coupling as $y_t = 3 y_1 y_2 y_3 / \sqrt{(y_1^2+y_2^2)(y_1^2+y_3^2)}$, reduces the number of free parameters by one. This allows us to rewrite the three Yukawa couplings, $y_1$, $y_2$, $y_3$ in terms of the top quark Yukawa and two mixing angles, $\tan\theta_{12}\equiv y_1/y_2$ and $\tan\theta_{13}\equiv y_1/y_3$, as defined in Eq.~\ref{eq:yukawa}.
\begin{eqnarray*}
y_1 &=& \frac{y_t}{3 \cos\theta_{12}\cos\theta_{13}} 
\end{eqnarray*}
\vspace{-4mm}
\begin{eqnarray}
y_2 &=& \frac{y_t}{3 \sin\theta_{12}\cos\theta_{13}}
\label{eq:yukawa} 
\end{eqnarray}
\vspace{-4mm}
\begin{eqnarray*}
y_3 &=& \frac{y_t}{3 \cos\theta_{12}\sin\theta_{13}}\cr
\end{eqnarray*}

\begin{figure}[tbh]
   \begin{center}
   \leavevmode
   \mbox{}
   \epsfxsize=8.5cm
   \epsffile{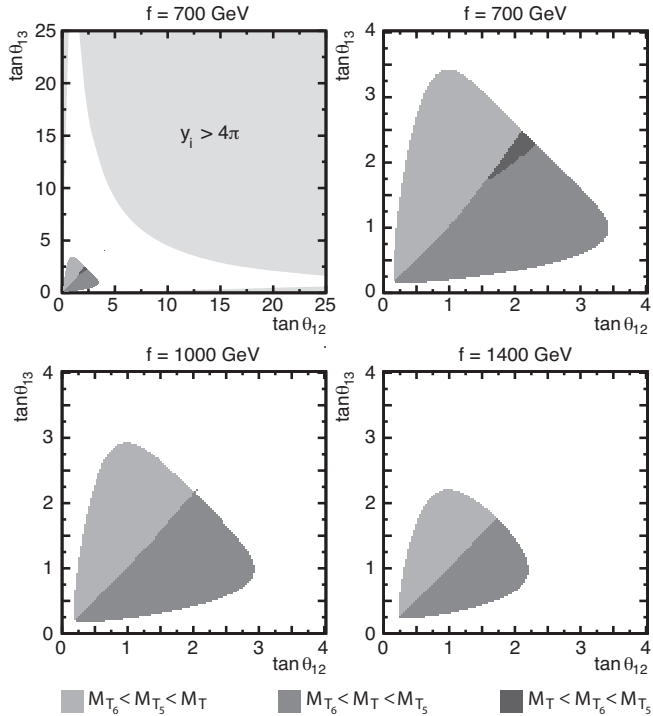}
   \end{center}
   \vspace{-7mm}
   \caption{Regions of $(\theta_{12},\theta_{13})$ allowed by theoretical constraints, for different values of scale $f$. The hierarchy of the heavy top-partners is shown in the legend at the bottom. The light grey region in the first plot shows the region of parameter space ruled out by the limit of $y_i < 4\pi$, ($i=1,2,3,t$), which is independent of the value of $f$. The remaining white region is ruled out by the fine-tuning constraint of $M_{T_i} < 2$~TeV, as discussed in Sec.~\ref{sec:constraint}.}
\label{fig:limits}
\end{figure}

\begin{figure}[tbh]
   \begin{center}
   \leavevmode
   \mbox{}
   \epsfxsize=8.5cm
   \epsffile{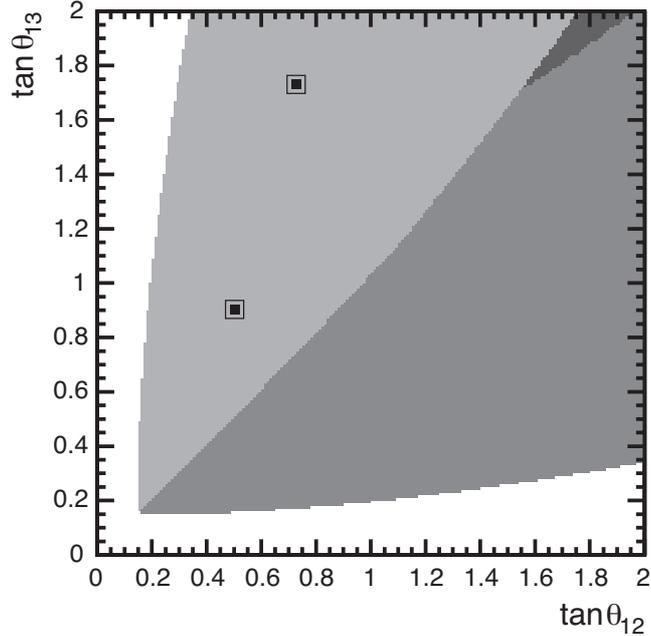}
   \end{center}
   \vspace{-7mm}
   \caption{Enlarged view of parameter space showing placement of isolated and non-isolated scenarios, for $f=700$~GeV. As evident from Fig.~\ref{fig:limits}, this region of ($\theta_{12}$, $\theta_{13}$) parameter space allows the greatest variability on the value of $f$ without violating any of the previously stated constraints.}
\label{fig:hierarchy}
\end{figure}


\begin{figure*}[tbhp]
   \begin{center}
   \leavevmode
   \mbox{}
   \epsfxsize=15cm
   \epsffile{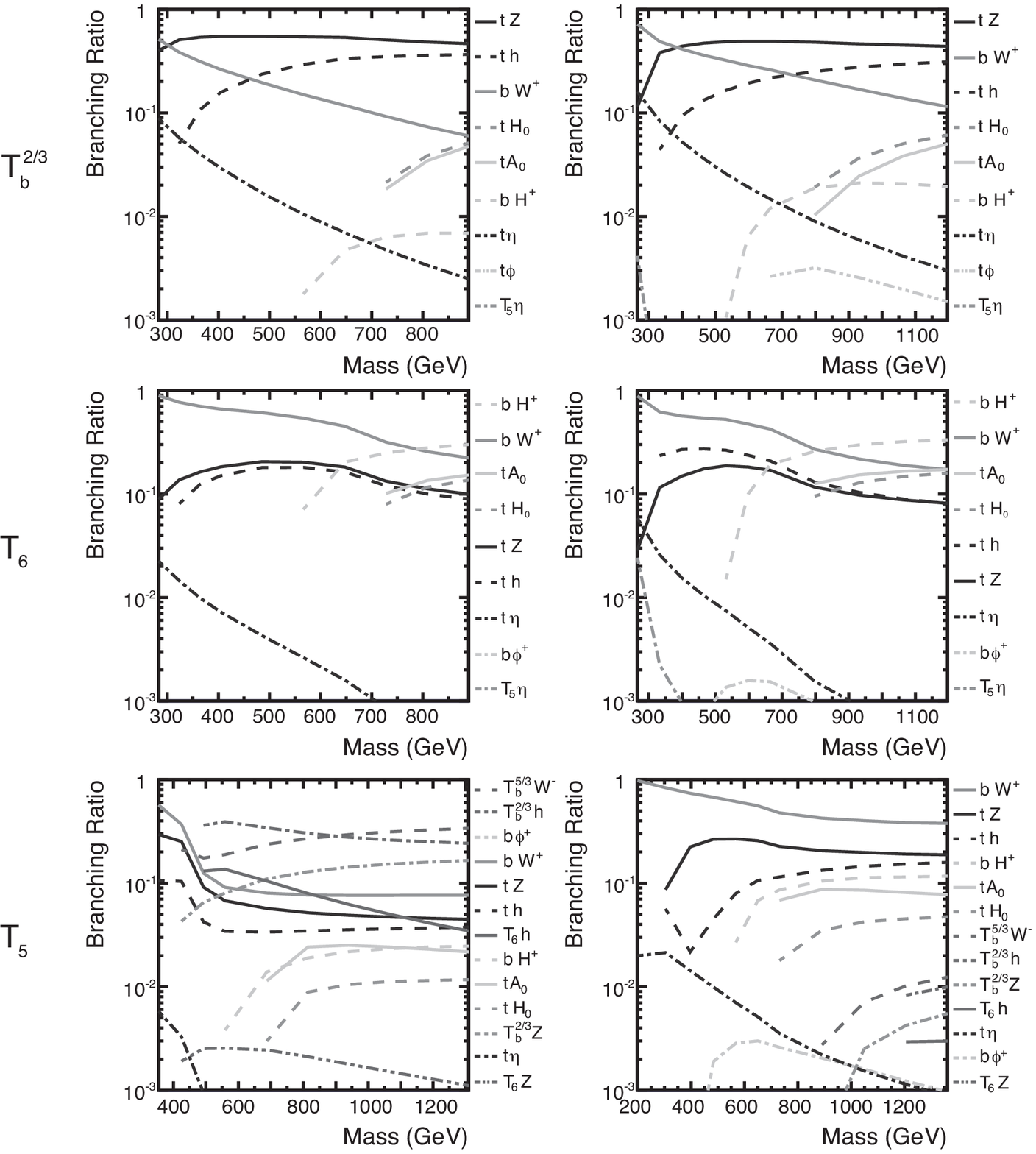}
   \end{center}
   \vspace{-7mm}
   \caption{Branching ratios for all two-body final states with $BR>0.1\%$. Branching ratios in the isolated scenario form the left column of plots, and the non-isolated scenario form the right column.}
\vspace{4mm}
\label{fig:br}
\end{figure*}


\section{Constraints}
\label{sec:constraint}

A number of theoretical constraints on the heavy fermion masses exist. Heavy fermion loop contributions to the Higgs potential and the lower bound on the Higgs boson mass from LEP suggest that $\tan\beta \gtrsim 1$~\cite{Schmaltz:2010ac}. Avoiding fine tuning in the top sector gives an upper bound on the heavy quark masses of 1-2 TeV. Lastly, constraints on the Yukawa couplings from perturbativity~\cite{Rathsman:2011yv, Altmannshofer:2010zt} set an upper bound of $4\pi$ on $y_1$, $y_2$, $y_3$ and $y_t$, which restricts the available region of $\tan\theta_{12}$ and $\tan\theta_{13}$ as shown by the large light grey region in the top left plot of Fig.~\ref{fig:limits}.

Scanning over all possible values of $\theta_{12}$, $\theta_{13}$ and $\tan\beta$, we find an upper constraint on the parameter $f$ in our diagonalization scheme of $f \lesssim 3100$~GeV when accounting for limits on $\tan\beta$ and fine tuning constraints on the heavy quark masses. Perturbativity requires that $f > v_{EW}$. However, the lower limit on $f$ can be set by experimental constraints from the production of heavy quarks, and varies significantly depending on the value of $\theta_{12}$ and $\theta_{13}$.

Figure~\ref{fig:limits} displays the mass hierarchy of the heavy top-like quarks in the region that satisfies the combined theoretical limits from perturbativity of the Yukawa couplings and from fine tuning from the heavy quark masses, for a given value of $\tan\beta = \sqrt{3}$ and for several values of the parameter $f$. For these plots, we expanded the mass matrix up to order $(v_{EW}/f)^4$ and used numerical diagonalization. For large values of $f$, the numerical values match the analytic values of Eq.~\ref{eq:Qmass}.
For large values of $\tan\theta_{1i}$ ($i = 2,3$), the value of the heaviest mass is significantly increased. This pushes either the $T_5$ or $T$ state to be significantly more massive than the others, and thus the available region of parameter space shrinks rapidly as the value of $f$ increases.

Since the full range of parameter space for this model is quite large, we restrict our analysis to two sample regions of parameter space that characterize the range of mass splittings for the heavy quarks. Due to the approximate $M_{T_5} \leftrightarrow M_T$ hierarchy symmetry about the central line of $\tan\theta_{12} = \tan\theta_{13}$, we only examine parameters on one side of the parameter space. The two regions of ($\theta_{12}$, $\theta_{13}$) that we consider are ($\pi/5$, $\pi/3$) and ($\tan^{-1}(0.5)$, $\tan^{-1}(0.9)$), which are referred to as the non-isolated and isolated scenarios, respectively, shown in Fig.~\ref{fig:hierarchy}. In the isolated scenario, the mass splitting between $T_5$ and $T_6$ is larger than in the non-isolated scenario. This increases the available decay modes for the $T_5$ state through cascade decays and reduces the branching fraction to the final states we consider. The branching ratios for the $T_6$, $T_5$ and $T_b^{2/3}$ are shown in Fig.~\ref{fig:br} for different masses of the heavy quarks in the two scenarios considered.

In both scenarios, the mass hierarchy for the heavy top partners is $M_{T_6}=M_{T_b^{2/3}} < M_{T_5} < M_{T}$. However, the mass of the $T$ quark is much heavier than the other masses and does not significantly contribute to the detector signatures for discovery.
 
For the calculation of the discovery limits, we use $\tan\beta = \sqrt{3}$, $M_h = 120$~GeV, $M_{A_0, H^\pm} = 500$~GeV, and $M_{\eta_0} = 10$~GeV. Masses of the scalar states primarily affect the point at which the decay mode becomes kinematically accessible. For the $A_0$, $H^\pm$ states, the mass is expected to be large enough that it does not significantly affect our results, while the branching fraction to the $\eta_0$ state is too small for the mass parameter to be significant. The mass of the $h$ state may affect the branching ratios of the heavy quarks in the mass region we are interested in, but recent results from ATLAS and CMS experiments suggest that a Higgs boson mass of 120~GeV is not unreasonable~\cite{CMS:2011h,ATLAS:2011h}. Small changes to the mass of the Higgs boson has a negligible effect on our results.

Further constraints on the values of $f$, $\tan\beta$, $\theta_{12}$ and $\theta_{13}$ may arise from loop induced precision electroweak measurements, including deviations to the SM $S$ and $T$ parameters, as well as from $K^0 - \bar{K}^0$, $B^0 - \bar{B}^0$ and $B_{s} - \bar{B}_{s}$ mixing.  Tree level contributions to both the $S$ and $T$ parameters vanish in the Bestest Little Higgs model, and therefore the model is not strongly constrained by precision electroweak measurements \cite{Schmaltz:2010ac}.  Since the $B$ quark does not mix with the $b$ quark, no limits should be found from $D^0 - \bar{D}^0$ mixing.  The couplings of the $T_5 - b - W$ and $T_6 - b - W$ vertices arise at order $v/f$, prior to implementing the CKM matrix, while the $T - b - W$ and $T_b^{2/3} - b - W$ vertices arise at order $v^2/f^2$.  Therefore, the primary additional contributions to meson mixing come from box diagrams involving exchanges of the $T_5$ and $T_6$ heavy quarks, which are suppressed by powers of $v/f$ relative to the SM contributions.  However, the scope of this study is direct heavy quark production and we refer interested readers to review similar studies in Ref.~\cite{ewprecision}.

\vspace{5mm}
\section{Pair Production \& Discovery Limits}
\label{sec:discovery_limits}

CMS recently presented discovery limits for pair production of a heavy top-like quark at the LHC running at $\sqrt{s} = 7$~TeV, assuming a 100\% branching ratio to the $bW$ final state~\cite{CMS:2011xo} with an integrated luminosity of 1.14~fb$^{-1}$, and a 100\% branching ratio to the $tZ$ final state~\cite{Chatrchyan:2011ay} with an integrated luminosity of 1.14~fb$^{-1}$ (previous study used 191~pb$^{-1}$~\cite{CMS:2011xo2}). 

We present and compare the results of the total contribution to the $b\bar{b}W^+W^-$ and $t\bar{t}ZZ$ cross sections from the pair production of the lightest heavy top-partner states ($T_6$, $T_5$ and $T_b^{2/3}$) in the Bestest Little Higgs model, given by:
\begin{equation}
\sigma_{tot} = \sigma_{T_5} + \sigma_{T_6}+\sigma_{T_b^{2/3}}
\label{eq:xsecbr}
\end{equation}
\begin{eqnarray*}
\sigma_{T_5}&=&\sigma(M_{T_5}) BR(T_5 \rightarrow qV)^2 \cr\cr
\sigma_{T_6}&=& \sigma(M_{T_6}) BR(T_6 \rightarrow qV)^2 \cr\cr
\sigma_{T_b^{2/3}}&=& \sigma(M_{T_b^{2/3}}) BR(T_b^{2/3} \rightarrow qV)^2
\end{eqnarray*}
where $q = t,b$ and $V=Z,W^+$ for the two final states considered. We found that the cross section for the $T$ state drops off rapidly due to the much larger mass, and therefore it is not included due to the very small contribution to the two channels considered here. Our cross sections are calculated using a fit of the NLO heavy quark production cross section data listed in Table V from Berger and Cao~\cite{Berger:2009qy} for $\sqrt{s} = 7$~TeV, which accounts for differences in the parameterization for lower masses of the heavy top partner, similar to the CMS analysis. The cross sections in ~\cite{Berger:2009qy} are found using the CTEQ6.6M PDF set~\cite{Stump:2003yu}. The branching ratios were calculated using the BRIDGE package~\cite{Meade:2007js}, and are plotted in Fig.~\ref{fig:br}. 

Since we combine cross section results from multiple quark states, we account for the differences in the kinematics for the cross section limits ($L_c$) by employing a weighted average, given in Eq.~\ref{eq:limit}, where $L(M)$ represents the limit on the cross section from the CMS results at the given heavy quark mass $M$.
\begin{eqnarray}
L_c(M_\mathrm{min})=&
\displaystyle\frac{ \sigma_{T_6} }{ \sigma_{tot} } L(M_{T_6})+\displaystyle\frac{\sigma_{T_b^{2/3}} }{ \sigma_{tot} } L(M_{T_b^{2/3}}) + \displaystyle\frac{\sigma_{T_5}}{\sigma_{tot}} L(M_{T_5})
\label{eq:limit}
\end{eqnarray}

Since the mass of the $T_6$ and $T_b^{2/3}$ quarks are identical in our expansion, the first two limits on the right hand side of Eq.~\ref{eq:limit} are identical. Additionally, the pair production cross section drops rapidly with the increasing mass of the heavy quark, and so $L_c$ is approximately the limit corresponding to the mass of the lightest state ($L_c$~$\approx$~$L(M_{T_6})$). The numerical results for the combined cross sections given in Eq.~\ref{eq:xsecbr}, as well as the masses and branching ratios for  the $T_6$, $T_b^{2/3}$ and $T_5$ quarks, are listed in Tables~\ref{table:bWdata} and \ref{table:tZdata}.

\begin{figure}[tbh]
   \begin{center}
   \leavevmode
   \mbox{}
   \epsfxsize=8.5cm
   \epsffile{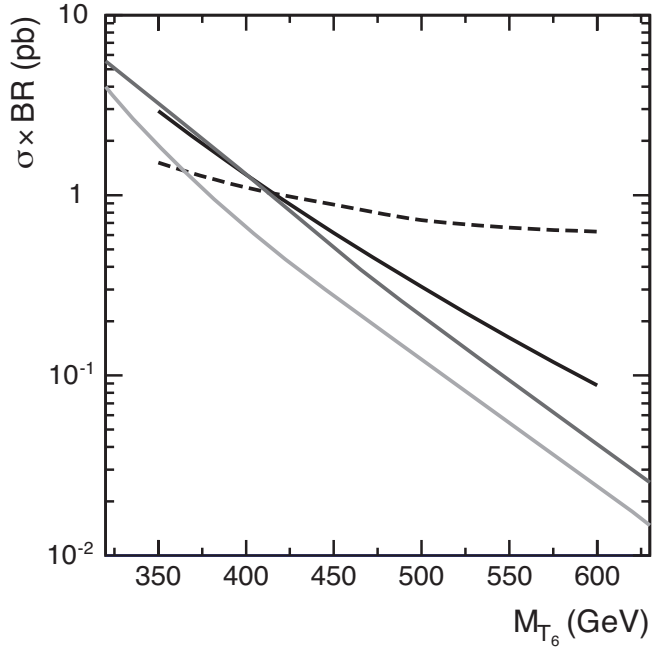}
   \end{center}
   \vspace{-7mm}
   \caption{Expected cross section times branching ratio (solid) and limits (dashed)~\cite{CMS:2011xo} in the $bW$ final state, for the isolated (light grey) and non-isolated (dark grey) scenarios. For comparison, the original heavy top-like quark pair production theoretical cross sections from~\cite{CMS:2011xo} are also included (black), which assumes a 100\% branching ratio to $bW$.}
\label{fig:xsec}
\end{figure}

\begin{figure}[tbh]
   \begin{center}
   \leavevmode
   \mbox{}
   \epsfxsize=8.5cm
   \epsffile{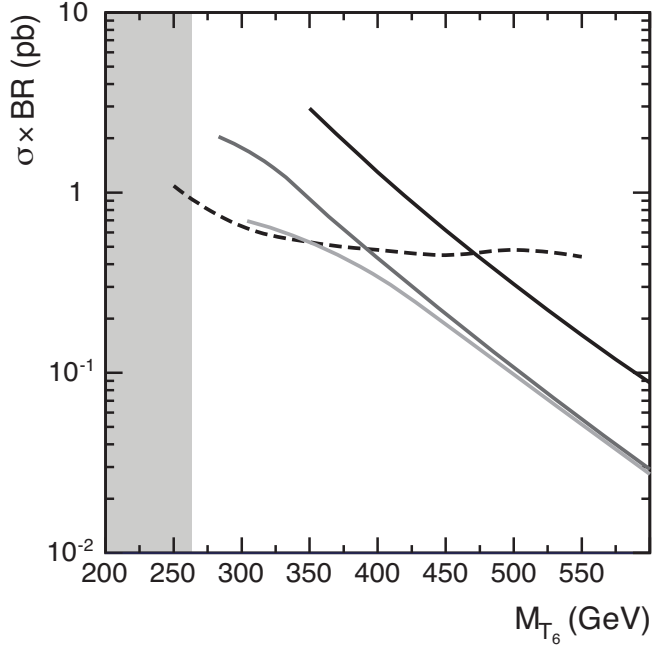}
   \end{center}
   \vspace{-7mm}
   \caption{Expected cross section times branching ratio (solid) and limits (dashed)~\cite{CMS:2011xo2} in the $tZ$ final state, for the isolated (light grey) and non-isolated (dark grey) scenarios. For comparison, the original pair production theoretical cross sections from~\cite{CMS:2011xo2} are also included (black), which assumes a 100\% branching ratio to $tZ$. The light grey region represents the region of phase space where the heavy quark mass is too small to produce an on-shell $tZ$ decay.}
\label{fig:xsecz}
\end{figure}

These results are plotted in Fig.~\ref{fig:xsec}, for the $bW$ channel, and Fig.~\ref{fig:xsecz}, for the $tZ$ channel, versus the mass of the lightest of the heavy top partners (the lightest state is $T_6$ over the range of the figures). Also plotted in these figures are the results presented by CMS for a model with $BR(T \rightarrow bW)=1$ (solid, black) for comparison, as well as the limits on the cross section presented in~\cite{CMS:2011xo, CMS:2011xo2}. The limits determined from Eq.~\ref{eq:limit} result in only a minor deviation from the cross section limits presented by CMS, and are omitted for visibility.

We find a lower mass limit for the degenerate $T_6$ and $T_b^{2/3}$ states of 364~GeV and 413~GeV for the isolated and non-isolated cases, respectively, from the $bW$ channel. For the $tZ$ channel, we find a lower mass limit of 347~GeV and 391~GeV, for the same two scenarios respectively. These mass limits correspond to a minimum value of $f$ of 892~GeV for the isolated scenario and 621~GeV for the non-isolated scenario, as calculated from the stronger of the discovery limits from the two decay modes.

The most recently published results from the ATLAS collaboration are for an integrated luminosity below 100~pb$^{-1}$, and are not considered in this study~\cite{ATLAS:2011}.\\

\vspace{5mm}
\section{Single Heavy Top Partner Production}

As first proposed in \cite{Han:2003qy}, heavy top partners can also be singly produced by t-channel $W$ exchange via the process $q b \rightarrow q^\prime T_i$ or by s-channel $W$ exchange via the process $q\bar{q}^\prime \rightarrow \bar{b} T_i$, where $T_i = T_5$, $T_6$, $T_b^{2/3}$, $T$ in the Bestest Little Higgs model.  Due to the presence of only one heavy top partner in the final state, and hence increased available phase space, the single production cross section falls off much more slowly than pair production as the top partner mass increases.  This is also due to the coupling of the heavy quark to the longitudinally polarized $W$ boson, which becomes enhanced at higher energies.  However, unlike QCD pair production, this is an electroweak process with a cross section that is more heavily influenced by the model parameters - in particular, the Yukawa mixing angles ($\theta_{12}$, $\theta_{13}$) of the heavy quark sector.

\begin{figure}[bh]
   \begin{center}
   \leavevmode
   \mbox{}
   \epsfxsize=8.5cm
   \epsffile{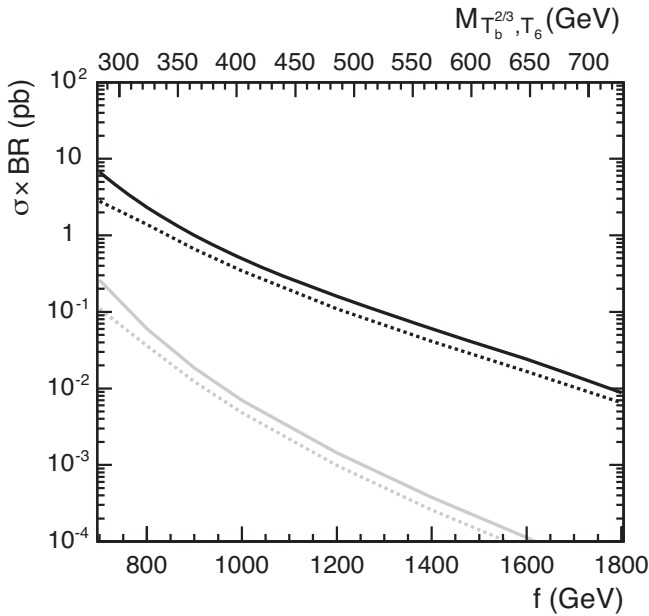}
   \end{center}
   \vspace{-7mm}
   \caption{Combined single production NLO cross section times branching ratio in the isolated scenario for $pp \rightarrow qb \rightarrow q^\prime T_i$ (black) and $pp \rightarrow q \bar{q}^\prime \rightarrow \bar{b} T_i$ (grey), for the $bW$ (solid) and $tZ$ (dashed) decay channels at $\sqrt{s} = 7 ~ \mathrm{TeV}$. Cross sections include the charge conjugate processes.}
\label{fig:single_i}
\end{figure}

\begin{figure}[tbh]
   \begin{center}
   \leavevmode
   \mbox{}
   \epsfxsize=8.5cm
   \epsffile{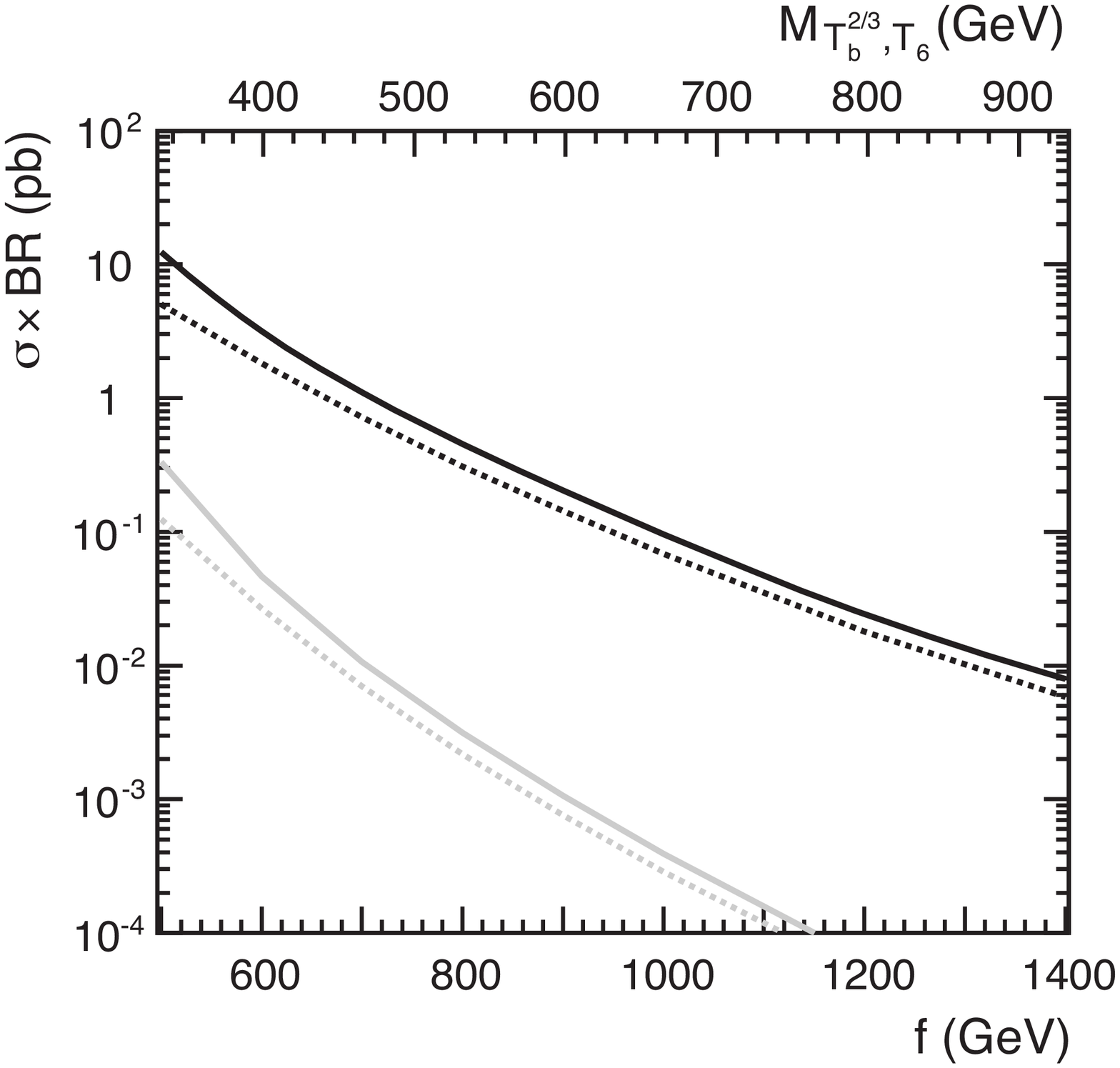}
   \end{center}
   \vspace{-7mm}
   \caption{Combined single production NLO cross section times branching ratio in the non-isolated scenario for $pp \rightarrow qb \rightarrow q^\prime T_i$ (black) and $pp \rightarrow q \bar{q}^\prime \rightarrow \bar{b} T_i$ (grey), for the $bW$ (solid) and $tZ$ (dashed) decay channels at $\sqrt{s} = 7 ~ \mathrm{TeV}$. Cross sections include the charge conjugate processes.}
\label{fig:single_ni}
\end{figure}

We determined the NLO single production cross sections of the heavy top partners using the function given in Eq. 1 in Berger and Cao~\cite{Berger:2009qy}, with the $A$, $B$, and $C$ parameters given in Table I of~\cite{Berger:2009qy} for $\sqrt{s} = 7$~TeV. For simplicity, the function given in \cite{Berger:2009qy} assumes that the $WT_{i}b$ vertices are the same as the Standard Model $Wtb$ vertex.  Our computed cross sections were therefore rescaled using the correct $WT_{i}b$ vertices of the Bestest Little Higgs model.  Our results were obtained using the CTEQ6.6M PDF set~\cite{Stump:2003yu}, and branching ratios of the heavy top partners to third generation fermions and Standard Model gauge bosons were calculated using BRIDGE~\cite{Meade:2007js}.  The resulting cross section for single production via t-channel $W$ exchange is given by, 
\begin{equation}
\sigma_{T_i} = \sigma(pp \rightarrow qb \rightarrow q^\prime T_i) BR(T_i \rightarrow q_3V),
\label{eq:xsecbr_Tj}
\end{equation}
where $q = u,c$; $q^\prime = d,s$; and $q_3 = t,b$ and $V=Z,W^+$ for the two final states considered.  This cross section was calculated for $T_i = T_5$, $T_6$, $T_b^{2/3}$.  The $T$ state is much heavier for the parameters considered, and was not included as its single production cross section is much smaller. 

We also include the charge-conjugate process  $q\bar{b} \rightarrow q^\prime \bar{T}_i$, which does not have the same cross section due to the different parton distribution functions of the initial-state quarks.  Its cross section is given by,
\begin{equation}
\sigma_{\bar{T}_i} = \sigma(pp \rightarrow q\bar{b} \rightarrow q^\prime \bar{T}_i) BR(\bar{T}_i \rightarrow \bar{q}_3 V),
\label{eq:xsecbr_Tbarj}
\end{equation}
where, in this case, $q = d,s$; $q^\prime = u,c$; and $\bar{q}_3 = \bar{t},\bar{b}$ and $V=Z,W^-$ for the two final states considered.  

The discovery cross section for single production is thus given by the sum of all contributions from heavy top-partners with similar detector signatures ($q^\prime q_3 V$ \& $q^\prime \bar{q}_3 V$), as in Eq.~\ref{eq:xsectot_Tj}.
\begin{equation}
\sigma_{single} = \sigma_{T_5} + \sigma_{T_6} + \sigma_{T_b^{2/3}} + \sigma_{\bar{T}_5} + \sigma_{\bar{T}_6} + \sigma_{\bar{T}_b^{2/3}}
\label{eq:xsectot_Tj}
\end{equation}

The cross section from Eq.~\ref{eq:xsectot_Tj} is plotted in Figs.~\ref{fig:single_i} and~\ref{fig:single_ni} versus $f$ (or alternatively, versus the mass of the lightest top partners) for the isolated and non-isolated scenarios respectively, for both the $bW$ and $tZ$ final states, again using $\tan\beta = \sqrt{3}$, $M_h = 120$~GeV, $M_{A_0, H^\pm} = 500$~GeV, and $M_{\eta_0} = 10$~GeV.  The isolated and non-isolated scenarios correspond to ($\theta_{12}$, $\theta_{13}$) values of ($\tan^{-1}(0.5)$, $\tan^{-1}(0.9)$) and ($\pi/5$, $\pi/3$), respectively.

Tables~\ref{table:bWdata} and~\ref{table:tZdata} present the numerical results for single production via t-channel $W$ exchange in the two scenarios, along with the cross sections for pair production.  Single production cross sections are generally larger than the pair production cross sections for larger top-partner masses, as expected. A further advantage of studying single production is that it does not suffer from the large $t\bar{t}$ backgrounds that exist for heavy top pair production.  A full analysis of backgrounds in each channel is beyond the scope of this paper and we leave this for a future study.

\begin{figure}[tbh]
   \begin{center}
   \leavevmode
   \mbox{}
   \epsfxsize=8.5cm
   \epsffile{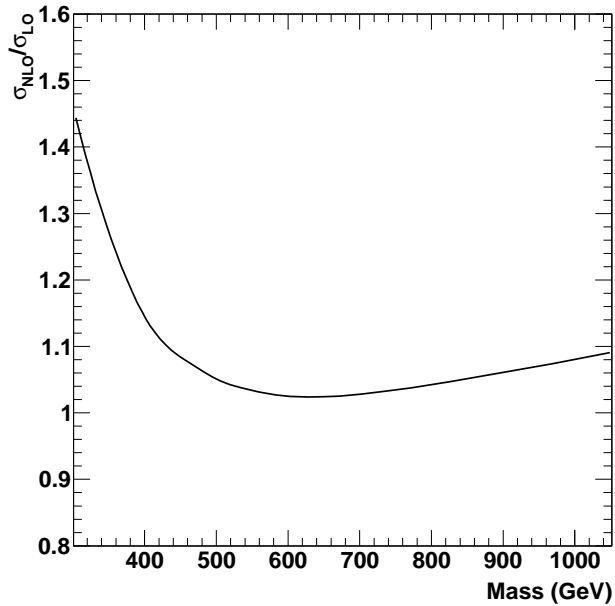}
   \end{center}
   \vspace{-7mm}
   \caption{Ratio of NLO and LO cross sections for single production via t-channel $W$ exchange for $\sqrt{s} = 7 ~ \mathrm{TeV}$, as a function of the top partner mass.  Only one curve is shown as the K factors were found to be identical for every scenario.}
\label{fig:single_kfactors}
\end{figure}

As a further check of our single production results, we calculated the LO cross sections for $pp \rightarrow qb \rightarrow q^\prime T_i$, where $T_i = T_5$, $T_6$, $T_b^{2/3}$, using MadGraph 5 v1.1.0~\cite{Alwall:2011uj} with the CTEQ6.6M PDF sets and compared these to the NLO cross sections calculated using the fit from~\cite{Berger:2009qy}.  The resulting K factors are shown in Fig.~\ref{fig:single_kfactors} and are similar to those found in~\cite{Campbell:2009gj}.  However, as explained in~\cite{Berger:2009qy}, the fit that was used to obtain the NLO cross sections is less valid at lower masses, resulting in a larger K factor in the low mass range.  One could improve the fit in the low mass range by increasing the number of terms in the polynomial expansion of the fitting function, as stated in~\cite{Berger:2009qy}, but this is beyond the scope of this paper since much of this low mass range has been ruled out based on the discovery limits calculated in section~\ref{sec:discovery_limits}.

The charge 2/3 heavy top partners can also be singly produced by s-channel $W$ exchange via the processes $q \bar{q}^\prime \rightarrow \bar{b} T_i$ and $q \bar{q}^\prime \rightarrow b \bar{T}_i$, where we consider $T_i = T_5$, $T_6$, $T_b^{2/3}$.  The NLO cross section for these processes was calculated in the same manner as the t-channel $W$ exchange process and is also plotted in Figs.~\ref{fig:single_i} and~\ref{fig:single_ni} for the isolated and non-isolated scenarios, respectively, for both the $bW$ and $tZ$ final states, again for $\sqrt{s} = 7$~TeV.  We find that the total cross section for $pp \rightarrow q \bar{q}^\prime \rightarrow \bar{b} T_i$ and $pp \rightarrow q \bar{q}^\prime \rightarrow b \bar{T}_i$ is roughly two orders of magnitude lower than the total cross section for $pp \rightarrow qb \rightarrow q^\prime T_i$ and $pp \rightarrow q\bar{b} \rightarrow q^\prime \bar{T}_i$.  This suggests that the t-channel $W$ exchange process is the most promising for studying single top partner production at the LHC.  However, a full analysis of the detector signatures and backgrounds in this channel is required to determine discovery capabilities in this channel.

ATLAS has recently measured a single top-quark production cross section of $90^{+32}_{-22}$ pb in the t-channel mode, after cuts, using $0.70~\mathrm{fb^{-1}}$ of data at $\sqrt{s} = 7~\mathrm{TeV}$~\cite{ATLAS:2011st}.  More luminosity is needed to reduce the uncertainty on this cross section and, since it is larger than the cross sections for single production of heavy top partners shown in Tables~\ref{table:bWdata} and \ref{table:tZdata}, harder cuts will be required to reduce this single top-quark background to a manageable level.

\section{Summary}

We presented a preliminary exploration of the heavy quark phenomenology in the Bestest Little Higgs model. In addition to exploring the regions of parameter space allowed by theoretical limits, we find the lower mass limits that can be set using recent CMS results for two scenarios of the parameters $\theta_{12}$ and $\theta_{13}$. We find a lower mass limit of 364~GeV for the isolated scenario and 413~GeV for the non-isolated scenario. In general, the limits in the Bestest Little Higgs model are lower than those given by CMS, which assume $BR(T\rightarrow bW)=1$ and $BR(T\rightarrow tZ)=1$, due to the lower branching ratios. It may be possible to enhance the search by combining results from multiple final states ($b\bar{b}W^+W^- + t\bar{t}ZZ + t\bar{b}W^-Z + b\bar{t}W^+Z$).  We also calculated the LO and NLO cross sections for single top partner production and determined that the cross sections are higher than that of pair production, particularly at heavy quark masses above approximately 500~GeV. We feel that these channels are worthwhile for the LHC experiments to explore further.
\acknowledgments
The authors would like to thank E. Berger, Q.-H. Cao, C. Hill, Y. Tu, M. Schmaltz, and J. Thaler  for helpful discussions. This research was supported in part by the Natural Sciences and Engineering Research Council of Canada.



\begin{table*}[p]
\scriptsize
\begin{center}
\caption[ Heavy quark pair production cross section and branching ratios. ] 
{\setlength{\baselineskip}{0.5cm} Summary of heavy quark pair and single production cross sections and branching ratios, assuming decays to the $bW$ final state. Branching ratios are for the stated heavy quark to the $bW$ final state.}
\begin{tabular}{ l | l | l | l | l | l | l | l }
\hline
\hline
\multicolumn{7}{c}{\textbf{Non-Isolated}} \\
\hline
\textbf{$f$ (GeV)} & \textbf{$M_{T_6, T_b^{2/3}}$ (GeV)} & \textbf{$M_{T_5}$ (GeV)} & \textbf{BR($T_6$)} & \textbf{BR($T_b^{2/3}$)} & \textbf{BR($T_5$)} & \textbf{$\sigma_{pair}$ (pb)} & \textbf{$\sigma_{single}$ (pb)}\\
\hline
500	&	332.3	&	304.0	&	0.620	&	0.489	&	0.835	&	7.36		&	12.3\\
600	&	398.8	&	396.5	&	0.566	&	0.413	&	0.739	&	1.39		&	3.17\\
700	&	465.3	&	483.5	&	0.542	&	0.360	&	0.676	&	0.387	&	1.10\\
800	&	531.8	&	567.6	&	0.524	&	0.321	&	0.614	&	0.127	&	0.455\\
900	&	598.2	&	649.9	&	0.474	&	0.287	&	0.559	&	0.0427	&	0.204\\
1000&	664.7	&	730.9	&	0.423	&	0.260	&	0.478	&	0.0145	&	0.0960\\
1200&	797.6	&	890.8	&	0.270	&	0.208	&	0.424	&	0.00168	&	0.0242\\
1400&	930.6	&	1048.7	&	0.218	&	0.169	&	0.402	&	0.000276	&	0.00797\\
\hline
\hline
\multicolumn{7}{c}{\textbf{Isolated}} \\
\hline
\textbf{$f$ (GeV)} & \textbf{$M_{T_6, T_b^{2/3}}$ (GeV)} & \textbf{$M_{T_5}$ (GeV)} & \textbf{BR($T_6$)} & \textbf{BR($T_b^{2/3}$)} & \textbf{BR($T_5$)} & \textbf{$\sigma_{pair}$ (pb)} & \textbf{$\sigma_{single}$ (pb)}\\
\hline
700	&	283.1	&	354.4	&	0.889	&	0.512	&	0.569	&	12.2		&	6.67\\
800	&	323.5	&	424.8	&	0.768	&	0.385	&	0.370	&	3.63		&	2.33\\
900	&	364.0	&	492.5	&	0.704	&	0.314	&	0.125	&	1.38		&	1.00\\
1000	&	404.4	&	558.6	&	0.660	&	0.263	&	0.091	&	0.612	&	0.497\\
1200	&	485.3	&	687.6	&	0.611	&	0.196	&	0.080	&	0.156	&	0.160\\
1400&	566.2	&	814.2	&	0.541	&	0.150	&	0.077	&	0.0419	&	0.0601\\
1600&	647.1	&	939.3	&	0.451	&	0.118	&	0.076	&	0.0109	&	0.0240\\
1800&	728.0	&	1063.4	&	0.316	&	0.092	&	0.076	&	0.00216	&	0.00878\\
\end{tabular}
\label{table:bWdata}
\end{center}
\end{table*}

\begin{table*}[p]
\scriptsize
\begin{center}
\caption[ Heavy quark pair production cross section and branching ratios. ] 
{\setlength{\baselineskip}{0.5cm} Summary of heavy quark pair and single production cross sections and branching ratios, assuming decays to the $tZ$ final state. Branching ratios are for the stated heavy quark to the $tZ$ final state.}
\begin{tabular}{  l | l | l | l | l | l | l | l  }
\hline
\hline
\multicolumn{7}{c}{\textbf{Non-Isolated}} \\
\hline
\textbf{$f$ (GeV)} & \textbf{$M_{T_6, T_b^{2/3}}$ (GeV)} & \textbf{$M_{T_5}$ (GeV)} & \textbf{BR($T_6$)} & \textbf{BR($T_b^{2/3}$)} & \textbf{BR($T_5$)} & \textbf{$\sigma_{pair}$ (pb)} & \textbf{$\sigma_{single}$ (pb)}\\
\hline
400	&	265.9	&	198.5	&	0.029	&	0.115	&	0.000	&	0.224	&	7.01\\
500	&	332.3	&	304.0	&	0.115	&	0.382	&	0.087	&	0.695	&	5.03\\
600	&	398.8	&	396.5	&	0.149	&	0.440	&	0.224	&	0.354	&	1.82\\
700	&	465.3	&	483.5	&	0.175	&	0.471	&	0.266	&	0.153	&	0.718\\
800	&	531.8	&	567.6	&	0.187	&	0.486	&	0.268	&	0.0651	&	0.309\\
900	&	598.2	&	649.9	&	0.182	&	0.493	&	0.257	&	0.0280	&	0.143\\
1000&	664.7	&	730.9	&	0.170	&	0.492	&	0.227	&	0.0121	&	0.0685\\
1200&	797.6	&	890.8	&	0.116	&	0.480	&	0.206	&	0.00240	&	0.0180\\
1400&	930.6	&	1048.7	&	0.098	&	0.464	&	0.198	&	0.000529	&	0.00581\\
\hline
\hline
\multicolumn{7}{c}{\textbf{Isolated}} \\
\hline
\textbf{$f$ (GeV)} & \textbf{$M_{T_6, T_b^{2/3}}$ (GeV)} & \textbf{$M_{T_5}$ (GeV)} & \textbf{BR($T_6$)} & \textbf{BR($T_b^{2/3}$)} & \textbf{BR($T_5$)} & \textbf{$\sigma_{pair}$ (pb)} & \textbf{$\sigma_{single}$ (pb)}\\
\hline
600	&	242.7	&	279.4	&	0.000	&	0.000	&	0.101	&	0.119	&	0.0501\\
700	&	283.1	&	354.4	&	0.089	&	0.400	&	0.296	&	2.04		&	2.80\\
800	&	323.5	&	424.8	&	0.137	&	0.508	&	0.252	&	1.37		&	1.39\\
900	&	364.0	&	492.5	&	0.163	&	0.538	&	0.091	&	0.736	&	0.665\\
1000	&	404.4	&	558.6	&	0.183	&	0.550	&	0.068	&	0.407	&	0.344\\
1200	&	485.3	&	687.6	&	0.205	&	0.550	&	0.057	&	0.130	&	0.110\\
1400&	566.2	&	814.2	&	0.203	&	0.544	&	0.052	&	0.0448	&	0.0411\\
1600&	647.1	&	939.3	&	0.182	&	0.535	&	0.049	&	0.0160	&	0.0167\\
1800&	728.0	&	1063.4	&	0.133	&	0.506	&	0.047	&	0.00545	&	0.00648\\
\end{tabular}
\label{table:tZdata}
\end{center}
\end{table*}


\end{document}